\documentclass[12pt]{iopart}
\usepackage{bm}

\usepackage{epsfig}
\usepackage{graphicx,float}
\usepackage{cite}
\usepackage{iopams}
\usepackage{latexsym}
\def\ee{\end{equation}}
\def\ba{\begin{eqnarray}}
\def\ea{\end{eqnarray}}
\def\la{~\mbox{\raisebox{-.6ex}{$\stackrel{<}{\sim}$}}~}
\def\ga{~\mbox{\raisebox{-.6ex}{$\stackrel{>}{\sim}$}}~}
\def\bq{\begin{quote}}
\def\eq{\end{quote}}

 at 10truept

\newcommand{\fnl}{\ensuremath{f_{\mathrm{NL}}}}
\newcommand{\beq}{\begin{equation}}
\newcommand{\eeq}{\end{equation}}
\newcommand{\beqa}{\begin{eqnarray}}
\newcommand{\eeqa}{\end{eqnarray}}
\newcommand{\bea}{\begin{eqnarray}}
\newcommand{\eea}{\end{eqnarray}}
\newcommand{\p}{\partial}
\newcommand{\al}{\alpha}
 
 \newcommand{\ep}{\epsilon}

\newcommand{\lle}{\left<}
\newcommand{\rgr}{\right>}

\newcommand{\vect}[1]{\bm{\mathrm{{#1}}}}

\def\la{~\mbox{\raisebox{-.6ex}{$\stackrel{<}{\sim}$}}~}
\def\ga{~\mbox{\raisebox{-.6ex}{$\stackrel{>}{\sim}$}}~}
\def\lesssim{~\mbox{\raisebox{-.6ex}{$\stackrel{<}{\sim}$}}~}
\def\ga{~\mbox{\raisebox{-.6ex}{$\stackrel{>}{\sim}$}}~}
\def\ltap{\ \raise.3ex\hbox{$<$\kern-.75em\lower1ex\hbox{$\sim$}}\ }
\def\gtap{\ \raise.3ex\hbox{$>$\kern-.75em\lower1ex\hbox{$\sim$}}\ }
\def\gl{\ \raise.5ex\hbox{$>$}\kern-.8em\lower.5ex\hbox{$<$}\ }
\def\roughly#1{\raise.3ex\hbox{$#1$\kern-.75em\lower1ex\hbox{$\sim$}}}

\begin{document}

\thispagestyle{empty}
\begin{flushright}
{\tt CERN-PH-TH/2009-253}\\
\end{flushright}

\title{Non-Gaussianity from Axion Monodromy Inflation}
\vskip1.5cm
\author{Steen Hannestad$^{1}$\footnote{\tt sth@phys.au.dk}, Troels Haugb\o{}lle$^{1,2,3}$\footnote{\tt haugboel@nbi.ku.dk}, Philip R. Jarnhus$^{1}$\footnote{\tt pjarn@phys.au.dk},  and Martin S. Sloth$^{4}$\footnote{\tt martin.sloth@cern.ch}}

\address{\em ${}^{1}$ Department of Physics and Astronomy, University of Aarhus, DK-8000 Aarhus C, Denmark}
\address{\em ${}^{2}$ Niels Bohr Institute, University of Copenhagen, Juliane Maries Vej 30, 2100 K\o benhavn  \O, Denmark}
\address{\em ${}^{3}$ Niels Bohr International Academy, Blegedamsvej 17, 2100 K\o benhavn  \O, Denmark}
\address{\em ${}^{4}$CERN, Physics Department, Theory Unit, CH-1211 Geneva 23, Switzerland}

\begin{abstract}
We study the primordial non-Gaussinity predicted from simple models of inflation with a linear potential and superimposed oscillations. This generic form of the potential is predicted by the axion monodromy inflation model, that has recently been proposed as a possible realization of chaotic inflation in string theory, where the monodromy from wrapped branes extends the range of the closed string axions to beyond the Planck scale. The superimposed oscillations in the potential can lead to new signatures in the CMB spectrum and bispectrum. In particular the bispectrum will have a new distinct shape.
We calculate the power spectrum and bispectrum of curvature perturbations in the model, as well as make analytic estimates in various limiting cases.
From the numerical analysis we find that for a wide range of allowed parameters the model produces a feature in the bispectrum with $\fnl \sim 5-50$ or larger while the power spectrum is almost featureless. This model is therefore an example of a string-inspired inflationary model which is testable mainly through its non-Gaussian features. Finally we provide a simple analytic fitting formula for the bispectrum which is accurate to approximately 5 \% in all cases, and easily implementable in codes designed to provide non-Gaussian templates for CMB analyses.
\end{abstract}

\vfill \setcounter{page}{0} \setcounter{footnote}{0}
\newpage

\setcounter{equation}{0} \setcounter{footnote}{0}

\section{Introduction}

In recent years there has been an increasing interest in the primordial non-Gaussianity predicted from inflation. Single field slow-roll inflation generally predicts non-Gaussianities to be unobservably small with present day techniques \cite{Acquaviva:2002ud,Maldacena:2002vr,Seery:2006vu,Seery:2008ax}. However, in non-minimal models of inflation the non-Gaussian signatures can be observable and non-Gaussianity is therefore an interesting observational window into the complexity of inflation.

Both from an observational and theoretical point of view the effective model of inflation is still uncertain. Since we do not know the microscopic origin of the effective theory of inflation, then the simplest theory of single field slow-roll inflation seems preferred from an Occam's razor point of view. Thus, with no prejudice about the underlying microscopic theory of inflation, we do not expect large non-Gaussianities. On the other hand, when inflation is embedded into a fundamental theory like string theory, it typically carries with it more complicated structures, and therefore from a model building point of view more complicated models of inflation can be attractive.

The simplest possible model of inflation, which is a single field model of chaotic inflation with a monomial potential, requires the field values of the inflaton to be trans-Planckian. This means that higher dimensional operators has to be prohibited by some symmetry in order not to destroy the flatness of the potential. In Natural Inflation models~\cite{Freese:1990rb,ArkaniHamed:2003wu} this is achieved by introducing an approximate shift symmetry, and assuming that the inflaton is the pseudo-Nambu-Goldstone boson mode (axion) of the theory. It has generally been considered non-straightforward to realize this idea in string theory, because the field values would typically be restricted to sub-Plackian values. This has led to the proposal that one could extend the field range by having many axions \cite{Dimopoulos:2005ac}. Recently another proposal has appeared\footnote{Alternative proposals have also been promoted in \cite{Kim:2004rp,Kaloper:2008fb,Ross:2009hg}}, where the field range is extended by a so called monodromy mechanism \cite{Silverstein:2008sg,McAllister:2008hb}. While not generic in this model, the model can have interesting oscillatory features embedded on top of the linear inflaton potential, depending on the details of the compactification \cite{Flauger:2009ab}.

Oscillatory features in the inflaton potential has already been studied by Chen, Easther and Lim as a way of generating large non-Gaussianity by a kind of resonance effect~\cite{Chen:2008wn}. The axion monodromy inflation model is a concrete model, which could lead to an observational signature of this particular type of non-Gaussinity. Since the small oscillations in the potential will be very hard to detect directly in the power spectrum, the signature in the bispectrum may provide the best way of experimentally verify such a scenario.

 Non-Gaussianity typically manifests itself through a non-vanishing three-point correlation function of the curvature perturbation, and is generally quantised in terms of $f_{NL}$, which measures the deviation in the curvature perturbation $\zeta$ from a Gaussian distributed variable $\zeta_g$ through~\cite{Maldacena:2002vr}
 \begin{equation}
 \zeta=\zeta_g-\frac{3}{5}f_{NL}\zeta_g^2~.
 \label{eq:fnldef}
 \end{equation}

At present, there are only good observational constraints on a scale independent $f_{NL}$ ($f_{NL}^{local}$), which gives $-10 \la f_{NL}^{local} \la 70$ at 95\% C.L. \cite{Curto:2009pv,Rudjord:2009mh,Smith:2009jr,Komatsu:2008hk,Yadav:2007yy} (note the preference for $f_{\rm NL}^{\rm local} > 0$ which in some analyses is significant at more than 95\% C.L.), but in the present model $f_{NL}$ will be scale dependent, and it is not clear how the bounds on $f_{NL}^{local}$ translate into a bound on the $f_{NL}$ in the present model, but one presumingly loose some sensitivity by only constraining $f_{NL}^{local}$ experimentially.

In the axion monodromy model, there are also non-trivial features in the power spectrum, and in the usual definition of $f_{NL}$ the power spectrum can therefore introduce some spurious scale dependence, which is not coming from the three-point function. Though this effect is at less than the per cent level for the specific models considered in this paper, we discuss a suitable alternative.

In order to calculate quantitatively the bispectrum in the axion monodromy inflation model, we have developed a code for computing  non-Gaussianity in general models of single field inflation with a minimal kinetic term.

\section{Bispectra in single field inflation}

In the case of single field inflation with some generic potential, one can write the minimally coupled Einstein-Hilbert action as
\beq
S = \int \mathrm{d}^4 x \sqrt{g}\left[\frac{1}{2}R-\frac{1}{2}(\p \phi)^2-V(\phi)\right]~.
\eeq
At zeroth order, this leads to the ordinary background equation of motion for the scalar field
\beq
\phi''+ 2\frac{a'}{a}\phi'+\frac{1}{a^2}\frac{dV}{d\phi} = 0
\eeq
and the time evolution of the slow-roll background can typically be expanded in the small slow-roll parameters
\begin{eqnarray}
\ep  & \equiv -\frac{\dot H}{H^2}=\frac{1}{2}\frac{\dot\phi^2}{H^2}\\
\eta & \equiv \quad\!\!\frac{\dot\ep}{\ep H}=2\frac{\ddot\phi}{H\dot\phi}+2\epsilon
\end{eqnarray}
where $H\equiv \dot a/a$ is the Hubble parameter.

Since we want to calculate correlation functions of fluctuations around this background, it is convenient to apply the ADM formalism \cite{Arnowitt:1962hi}. Here the perturbed metric takes the form
\beq
ds^2= -N^2dt^2+h_{ij}(dx^i+N^idt)(dx^j+N^jdt)
\eeq
where the lapse and the shift $N, N^i$ acts as Langrange multipliers when the perturbed metric is inserted into the action. Only $\phi$ and $h_{ij}$ are dynamical variables, and in a convenient gauge that fix time and spatial reparametrizations, one can write
\beq
h_{ij} = a^2(t)\left[e^{2\zeta}\delta_{ij}+\gamma_{ij}\right]~,
\eeq
while having no fluctuations in the inflaton field. The physical degrees of freedom in this gauge are the scalar curvature perturbation $\zeta$ and the transverse and traceless tensor mode $\gamma_{ij}$. We can safely ignore tensor perturbation in this work, as any contribution from tensor perturbations to the scalar three-function only arise through loop diagrams.

Inserting the perturbed gauge fixed metric into the action, we can derive the action for the curvature perturbation iteratively to any order in perturbation theory. From the quadratic action of $\zeta$
\beq
S = \frac{1}{2}\int \mathrm{d} t \mathrm{d}^3 x \frac{\dot\phi^2}{H^2}\left[a^2\dot\zeta^2-a(\p\zeta)^2\right]~,
\eeq
upon quantizing the $\zeta$ field in the Bunch-Davies vacuum, one obtains the power spectrum, $P_{\zeta}(k)$, of curvature perturbations
\beq
\left<\zeta_{{\bf k}_1}\zeta_{{\bf k}_2}\right> \equiv (2\pi)^3\delta^3({\bf{k}}_1+{\bf{k}}_2)P_{\zeta}(k)~,
\eeq
which is related to the usual scale independent power spectrum by $P_{\zeta}(k)=(2\pi^2)/k^3 \mathcal{P}_{\zeta}(k)$.

Similarly one can define the bispectrum  $B_\zeta$ by
\begin{equation}
    \langle \zeta(\vect{k}_1)
        \zeta(\vect{k}_2)
        \zeta(\vect{k}_3) \rangle
    \equiv (2\pi)^3 \delta(\sum_a \vect{k}_a)
    B_\zeta(\vect{k}_1,\vect{k}_2,\vect{k}_3) ~,
\end{equation}
The strength of the non-Gaussian signal in the bispectrum can conveniently parameterised in terms of the scale independent non-linearity parameter $\fnl$, by defining
\begin{equation}
    \label{fnl}
    B_\zeta \equiv \frac{6}{5} \fnl
    [ P_{\zeta}(k_1) P_{\zeta}(k_2)+ \mbox{2 permutations} ]
    \,.
\end{equation}

Now, in order to calculate the three-point function one calculates the action for the curvature perturbations, $\zeta$, to third order\footnote{The fourth order action has been calculated in \cite{Sloth:2006az}, \cite{Seery:2006vu}, \cite{Jarnhus:2007ia}.
}, which, up to total derivatives, leads to the interaction Hamiltonian~\cite{Maldacena:2002vr}
\bea\label{Hi}
H_{int}(\tau) &=& -\int \mathrm{d}^3 x\left[ a\ep^2\zeta\zeta'^2+ a\ep^2\zeta(\p\zeta)^2-2\ep\zeta'(\p\zeta)(\p\chi)\right.\nonumber\\
& &\left. +\frac{a}{2}\ep\eta'\zeta^2\zeta'+\frac{\ep}{2a}(\p\zeta)(\p\chi)(\p^2\chi) +\frac{\ep}{4a}(\p^2\zeta)(\p\chi)^2\right]~,
\eea
where $\p^2\chi =\ep \dot\zeta$.

 One can, through this result, evaluate the three-point function of curvature perturbations after horizon exit, by means of the in-in formalism:
\beq\label{eq:bispecintegral}
\lle  \zeta_{k_1}(\tau)\zeta_{k_2}(\tau)\zeta_{k_3}(\tau)\rgr =-i\int_{\tau_0}^\tau \mathrm{d}\tau a\lle  [\zeta_{k_1}(\tau)\zeta_{k_2}(\tau)\zeta_{k_3}(\tau),H_{int}(\tau')]\rgr~,
\eeq
together with a field redefinition accounting for boundary terms that was neglected in eq.~(\ref{Hi})
\beq
\lle  \zeta_{k_1}(\tau)\zeta_{k_2}(\tau)\zeta_{k_3}(\tau)\rgr =\lle  \tilde\zeta_{k_1}(\tau) \tilde\zeta_{k_2}(\tau) \tilde\zeta_{k_3}(\tau)\rgr + \eta\lle   \tilde\zeta^2_{k_1}(\tau) \tilde\zeta_{k_2}(\tau) \tilde\zeta_{k_3}(\tau)\rgr +\textrm{sym.}~.
\eeq

In a scenario where there are periodic ripples in the inflaton potential, the slow-roll parameters will generally oscillate. In such a scenario, it was demonstrated by Chen, Easther and Lim~\cite{Chen:2008wn} that when the perturbations modes are inside the horizon and oscillate, they can interfere constructively with the oscillations in the couplings in the interaction Hamiltonian above, which are determined by the slow-roll parameters $\ep$, $\eta'$. As a consequence there can be a resonant amplification of non-Gaussianity from the bispectrum.

\section{The effective model of inflation from axion monodromy}

It has been demonstrated that in the large field limit, which is valid during inflation, the axion monodromy model is described by a linear potential for the axion $\phi$ and an axion action of the form \cite{McAllister:2008hb}
\beq
S_{\phi}=\int\mathrm{d}^4x\sqrt{g}\left(\frac{1}{2}(\partial\phi)^2-\mu^3\phi\right)+{\rm corrections}~.
\eeq
One can show that in order to obtain 60 e-folds of inflation with the right level of curvature perturbations in the leading order linear potential, one should have
\bea
\phi&\sim&11M_{{\rm pl}}\\
\mu&\sim&6 \times 10^{-4}M_{{\rm pl}}
\eea
as initial conditions. Though the brane-induced inflaton potential described above is the leading effect for breaking shift symmetries in the class of models considered here, there are other effect present as well. Among the more important is production of instantons, which, in this case, give rise to periodic corrections to the potential. One can therefore write down the effective potential containing these two contributions as \cite{McAllister:2008hb}
\beq
V_{eff} =\mu^3\phi (1+\al_1\cos(\phi/f))+\al_2 M_s^4\cos(\phi/f)~,
\label{eq:effpot}
\eeq	
where $M_s$ is the string scale and $\alpha_i$ are dimensionless parameters which are expected to be much smaller than one, $\al_i\ll1$, depending on the details of the string theory compactifications.

We will study the two cases $\al_1\ll \al_2$ and $\al_1\gg \al_2$, leading to two limiting potentials. For $\al_1\gg\al_2$
\beq
V^{(1)}_{eff} =\mu^3\phi (1+\al\cos(\phi/f)),
\eeq	
where we will assume $\al \equiv \al_1 \sim 10^{-5}$, such that effect on the power spectrum is at the per cent level and not yet excluded by data. The other case is $\al_1\ll\al_2$ where
\beq
V^{(2)}_{eff} =\mu^3\phi+\Lambda^4\cos(\phi/f),
\eeq	
and where we will assume $\Lambda^4 =\al_2 M_s^4$ is such that the effect on the power spectrum is at the per cent level.

The scalar power spectrum for the potential in the second model was computed analytically in~\cite{Flauger:2009ab}. We will return to this point later in Eq.~(\ref{eq:ps}).

\subsection{Isosceles limit estimates of the bispectrum}

As discussed in section two, one can write $f_{NL}$ as a combination of the three-point function of the Fourier transform of $\zeta$ and the power spectra of the curvature perturbations, yielding
\begin{equation}\label{fnl2}
f_{NL}=-\frac{10}{3}\frac{(k_1k_2k_3)^3}{\mathcal{P}(k_1)\mathcal{P}(k_2)k_3^3+{\rm 2 \, perms.}}\frac{\left<\zeta_{\bm{k}_1}\zeta_{\bm{k}_2}\zeta_{\bm{k}_3}\right>}{(2\pi)^7\delta^{(3)}\left(\bm{k}_1+\bm{k}_2+\bm{k}_3\right)}~,
\end{equation}
where $k_i$ is the magnitude of the momentum $\bm{k}_i$. For the case of our models the fluctuations in the power spectrum is of the order one per cent. However if one should concern oneself with models with larger fluctuations in the power spectrum, one may benefit from using a similar definition, given by
\begin{equation}
\tilde f_{NL}=-\frac{10}{3}\frac{(k_1k_2k_3)^3}{\tilde{\mathcal{P}}^2 (k_1^3+k_2^3+k_3^3)}\frac{\left<\zeta_{\bm{k}_1}\zeta_{\bm{k}_2}\zeta_{\bm{k}_3}\right>}{(2\pi)^7\delta^{(3)}\left(\bm{k}_1+\bm{k}_2+\bm{k}_3\right)}~,
\end{equation}
where $\tilde{\mathcal{P}}$ is approximately the amplitude of the power spectrum at the $k$'s considered. It should be noted that this definition reduces to the $\frac{\mathcal{G}}{k_1k_2k_3}$ presented in~\cite{Chen:2008wn} in the equilateral limit, but differs in the squeezed limit, where $\frac{\mathcal{G}}{k_1k_2k_3}$ grows as $\frac{k_1}{k_3}$ for triangles of the type $k_1=k_2\gg k_3$.

We proceed to find an estimate for the case $k_1=k_2\geq k_3$. Though such an estimate does not exist in general, we can obtain one by interpolating between the equilateral limit ($k_1=k_2=k_3$) and the squeezed limit ($k_1=k_2\gg k_3$).
\subsubsection{Equilateral Limit}
The model is essentially a linear slow-roll potential with a small oscillating perturbation. We can therefore follow the approach of Chen, Easther and Lim \cite{Chen:2008wn} and calculate the first order effect of the small oscillating term as a small correction. In this way one has $\ep = \ep_0 + \ep_{osc}$,  $\eta = \eta_0 + \eta_{osc}$. As the background solution (subscript 0), one can start from
\beq
3H\dot\phi +\mu^3 =0
\eeq
to obtain
\beq
\phi_0=\frac{1}{2^{2/3}}(-\sqrt{3}\mu^{3/2}t+2\phi_i^{3/2})^{2/3}
\eeq
and
\beq
\ep_0 = \frac{1}{2\phi_0^2}~,\qquad \eta_0 = \frac{2}{\phi_0^2}~.
\eeq

As argued in~\cite{Chen:2008wn} one can estimate $f_{NL}$ in the equilateral limit by
\beq
\fnl^{(eq)}\sim-f_A^{(eq)}\sin\left(\frac{\ln K}{\phi f} + \textrm{phase}\right)
\label{eq:equiestimate}
\eeq
in the equilateral limit. Here $K=k_1+k_2+k_3$. The factor 2 missing, compared to~\cite{Chen:2008wn}, is due to the underlying linear potential here, as opposed to a quadratic in~\cite{Chen:2008wn}. The resonance amplitude can be estimated as
\beq
f_A^{(eq)}\sim \frac{10}{9}\frac{\dot\eta_A}{H\sqrt{f\phi}}~,
\eeq
where $\dot\eta_A$ is the amplitude of $\dot\eta_{osc}$, which can be found from\\[1ex]
\hspace*{-1cm}
\begin{minipage}[t]{0.52\textwidth}
\begin{eqnarray}
\hspace*{-1cm} \ep_{osc} & = & \frac{\dot H_{osc}}{\dot H_0}\ep_0 = -3\frac{\Lambda^4}{\mu^3}\frac{1}{\phi_0}\cos\left(\frac{\phi_0}{f}\right)\\
\hspace*{-1cm} \eta_{osc} &= & \frac{\dot\ep_{osc}}{\dot\ep_0}\eta_0 = -6\frac{\Lambda^4}{\mu^3f}\sin\left(\frac{\phi_0}{f}\right)
\end{eqnarray}
\begin{center}
$\Lambda^4$-model
\end{center}
\vspace{1ex}
\end{minipage}
\hfill\vrule\hfill
\begin{minipage}[t]{0.52\textwidth}
\begin{eqnarray}
\hspace*{-1cm} \ep_{osc} & = & \frac{\dot H_{osc}}{\dot H_0}\ep_0 = -3\alpha\cos\left(\frac{\phi_0}{f}\right)\\
\hspace*{-1cm} \eta_{osc}& = & \frac{\dot\ep_{osc}}{\dot\ep_0}\eta_0  = -6\frac{\alpha\phi_0}{f}\sin\left(\frac{\phi_0}{f}\right)
\end{eqnarray}
\begin{center}
$\alpha$-model
\end{center}
\vspace{1ex}
\end{minipage}	\\[1ex]
We therefore find the resonance amplitude to be\\[1ex]
\hspace*{-1cm} \begin{minipage}[t]{0.52\textwidth}
\begin{equation}
\hspace*{-1cm} f_A^{(eq)}\sim\frac{10}{9}\frac{\Lambda^4}{\phi\mu^3}\frac{1}{f^{5/2}\phi^{1/2}}
\end{equation}
\begin{center}
$\Lambda^4$-model
\end{center}
\vspace{1ex}
\end{minipage}
\hfill\vrule\hfill
\begin{minipage}[t]{0.52\textwidth}
\begin{equation}
\hspace*{-1cm} f_A^{(eq)}\sim\frac{10}{9}\alpha\frac{1}{f^{5/2}\phi^{1/2}}
\end{equation}
\begin{center}
$\alpha$-model
\end{center}
\vspace{1ex}
\end{minipage}	\\[1ex]

\subsubsection{Squeezed limit}

Let us now consider the different limit $k_1=k_2\gg k_3$  (for simplicity we define $k_1=k_2=k$ and $k_3=m$).  In this limit the long wavelength mode of $\zeta_{\vect{k}_3}$, will act as a constant rescaling of the background of the two shorter wavelength modes. As shown by Maldacena~\cite{Maldacena:2002vr}, one can thus calculate the three point correlation function in this limit by calculating the correlation of the long wavelength mode with the variation of the two-point function of the short wavelength modes on the background of the long wavelength mode \cite{Maldacena:2002vr,Creminelli:2004yq}
\beq
\lim_{k_3\to 0}\left< \zeta_{\vect{k}_1}\zeta_{\vect{k}_2}\zeta_{\vect{k}_3}\right> = \left<\zeta_{\vect{k}_3}\left<\zeta_{\vect{k}_1}\zeta_{\vect{k}_2}\right>_B\right>~.
\eeq
Then by a Taylor expansion of $\left<\zeta_{\vect{k}_1}\zeta_{\vect{k}_2}\right>_B$ on the unperturbed background, and using $\p/\p\zeta\to k\p/\p k$ (since the effect of $\zeta$ is to take $k \to k-k\zeta$), one finds \cite{Maldacena:2002vr,Creminelli:2004yq}
\bea
\lim_{k_3\to 0}\left< \zeta_{\vect{k}_1}\zeta_{\vect{k}_2}\zeta_{\vect{k}_3}\right>& =& -(2\pi)^3\delta^3(\vect{k}_1+\vect{k}_2+\vect{k}_3)P_{\zeta}(k_3)\frac{2\pi^2}{k^3}\frac{d}{d\ln(k) } \mathcal{P}_{\zeta}(k)\nonumber\\
&=&-(n_s-1)(2\pi)^3\delta^3(\vect{k}_1+\vect{k}_2+\vect{k}_3)P_{\zeta}(k_3)P_{\zeta}(k)
\eea
which can be compared to eq.(\ref{fnl}), and one can read off $f_{NL}^{(sq)}=(5/12)(n_s-1)$. Here $n_s$ is the spectral index.

As it was shown in~\cite{Flauger:2009ab} the scalar power spectrum for the potential in the second model can be written in the form
\begin{equation}\label{eq:ps}
\mathcal{P}_\zeta(k)=\mathcal{P}_\zeta(k_*)\left(\frac{k}{k_*}\right)^{\tilde n_s-1}\left[1+\delta n_s\cos\left(\frac{\phi_k}{f}\right)\right]~.
\end{equation}
with the short hand notation of~\cite{Flauger:2009ab}:\\[1ex]
\hspace*{-1cm} \begin{minipage}[t]{0.52\textwidth}
\begin{equation}
\hspace*{-1cm} \delta n_s=\frac{12\Lambda^4}{f\mu^3}\frac{\sqrt{\frac{\pi}{8}\coth\left(\frac{\pi}{2f\phi_*}\right)f\phi_*}}{\sqrt{1+(3f \phi_*)^2}}
\end{equation}
\begin{center}
$\Lambda^4$-model
\end{center}
\vspace{1ex}
\end{minipage}
\hfill\vrule\hfill
\begin{minipage}[t]{0.52\textwidth}
\begin{equation}
\hspace*{-1cm} \delta n_s=\frac{12\alpha\phi_*}{f}\frac{\sqrt{\frac{\pi}{8}\coth\left(\frac{\pi}{2f\phi_*}\right)f\phi_*}}{\sqrt{1+(3f \phi_*)^2}}
\end{equation}
\begin{center}
$\alpha$-model
\end{center}
\vspace{1ex}
\end{minipage}	\\[1ex]
The $\phi_k$ is the field value, when the momentum $k$ crosses the horizon.
\begin{equation}
\phi_k=\sqrt{\phi_*^2-2\ln\left(\frac{k}{k_*}\right)}\simeq\phi_*-\frac{\ln\left(\frac{k}{k_*}\right)}{\phi_*}~.
\end{equation}
The starred quantities are merely an arbitrarily chosen fix point, where $k_*$ is the momentum that crosses the horizon at $\phi=\phi_*$. Though complicated, the expression for $\delta n_s$ reduces to\\[1ex]
\hspace*{-1cm} \begin{minipage}[t]{0.52\textwidth}
\begin{equation}
\hspace*{-1cm} \delta n_s\approx6\frac{\sqrt{\frac{\pi}{2}}\Lambda^4\phi_*^{1/2}}{f^{1/2}\mu^3}
\end{equation}
\begin{center}
$\Lambda^4$-model
\end{center}
\vspace{1ex}
\end{minipage}
\hfill\vrule\hfill
\begin{minipage}[t]{0.52 \textwidth}
\begin{equation}
\hspace*{-1cm} \delta n_s\approx6\frac{\sqrt{\frac{\pi}{2}}\alpha\phi_*^{3/2}}{f^{1/2}}
\end{equation}
\begin{center}
$\alpha$-model
\end{center}
\vspace{1ex}
\end{minipage}	\\[1ex]
for $f\ll1$. While the expression for $\delta n_s$ is very precise for $f \gtrsim 10^{-3}$ it overestimates significantly the correct value for $f \lesssim 10^{-4}$.

As $\delta n_s\ll 1$ we find
\begin{equation}
n_s-1=\tilde{n}_s-1+\frac{\delta n_s}{f\phi_*}\sin\left(\frac{\phi_{k}}{f}\right)
\end{equation}
to the first order in $\delta n_s$. With this one can find an expression for $f_{NL}$ in the squeezed limit:
\begin{equation}
\fnl^{(sq)}\sim f_A^{(sq)}\sin\left(\frac{\phi_k}{f}\right)~,~f_A^{(sq)}\equiv\frac{5}{12}\frac{\delta n_s}{f\phi_*}
\end{equation}
to the first order in $\delta n_s$ for $k_1=k_2\gg k_3$.

This result can now be combined with eq. (\ref{eq:equiestimate}) to form a general estimate for isosceles momentum triangles ($k_1=k_2\geq k_3$):
\begin{equation} \label{fnlest}
f_{NL}\sim\frac{5}{12}(\tilde n_s-1)+\left[f_A^{(sq)}+(f_A^{(eq)}-f_A^{(sq)})\frac{m}{k}\right]\sin\left(\frac{\ln K}{\phi f}+\textrm{phase}\right)~,
\end{equation}
where for completeness we have included the slow-roll contribution from the underlying linear potential, as the first term.

\subsection{Semiclassical estimates of the trispectrum}

If the bispectrum can become large in a given model, one might be interested in the size of higher order correlation functions as well. While it is beyond the scope of this paper to study the trispectrum in details, we can still make some generic semiclassical estimates in certain limits. Most obviously we can study the trispectrum in the squeezed limit \cite{Seery:2006vu,Huang:2006eha}, but it is also possible to estimate a particular contribution to the non-Gassianity in the counter-collinear limit \cite{Seery:2008ax}.

\subsubsection{Squeezed limit}

One can imagine to make a definition of  $\tau_{NL}$ analogous to the definition of $f_{NL}$ in eq.~(\ref{fnl2}), relating it to the four-point function as
\begin{equation}
\tau_{NL}\sim\frac{1}{\tilde{\mathcal{P}}^3(2\pi)^9}\frac{(k_1k_2k_3k_4)^3}{\sum_ik_i^3}\frac{\left<\zeta(\bm{k}_1)\zeta(\bm{k}_2)\zeta(\bm{k}_3)\zeta(\bm{k}_4)\right>}{\delta^{(3)}\left(\sum_i\bm{k}_j\right)}
\end{equation}
By repeating the above calculation for $k\equiv k_1=k_2=k_3\gg k_4$ under the same assumptions, one can get an estimate of the behaviour of the trispectrum in the squeezed limit
\beq
\lim_{k_4 \to 0}\left< \zeta_{\vect{k}_1}\zeta_{\vect{k}_2}\zeta_{\vect{k}_3}\zeta_{\vect{k}_4}\right> = -(2\pi)^3\delta^3(\vect{k}_1+\vect{k}_2+\vect{k}_3+\vect{k}_4)P_{\zeta}(k_4)\frac{1}{k^6}\frac{d}{d\ln(k) } \left( k^6 B_{\zeta}(k)\right)~.
\eeq

In order to evaluate this expression approximately, we can insert the estimate for the equilateral bispectrum obtained in eq.~(\ref{eq:equiestimate}) to find
\begin{equation}
\tau_{NL}\sim\frac{3}{20}\fnl^{(eq)}\frac{\mathcal{P}(k)^2\mathcal{P}(k_4)}{\tilde{\mathcal{P}}^3}\left[2(n_s-1)+\frac{\tan\left(\frac{\ln(3k)}{f\phi_*}+\Phi\right)}{f\phi_*}\right]~.
\end{equation}
However, we can obtain an even simpler estimate of the size of the trispectrum in specific configurations, by considering the double squeezed limit, where $k\equiv k_1 = k_2 >> k_3 >> k_4$. Iteratively, we then obtain
\bea
\lim_{k_3, k_4 \to 0}\left< \zeta_{\vect{k}_1}\zeta_{\vect{k}_2}\zeta_{\vect{k}_3}\zeta_{\vect{k}_4}\right>\sim  (n_s-1)^2 (2\pi)^3\delta^3(\vect{k}_1+\vect{k}_2+\vect{k}_3+\vect{k}_4)P_{\zeta}(k_3)P_{\zeta}(k_4)P_{\zeta}(k)~.
\eea
This implies that in this limit $\tau_{NL} \sim (\fnl^{(sq)})^2$.

\subsubsection{Counter-collinear limit}

Let us consider the contribution to the four-point function from the exchange of a scalar mode between two pairs of external scalar modes. In the counter collinear limit where the momentum of the exchanged mode goes to zero $\vect{k}_1+\vect{k}_2 \to 0$, the contribution to the four-point function from the exchange process can be expressed as the correlation of a pair of two-point functions $\left<\zeta_{\vect{k}_1}\zeta_{\vect{k}_2}\right>$, $\left<\zeta_{\vect{k}_3}\zeta_{\vect{k}_4}\right>$ due to the presence of the long wavelength scalar mode \cite{Seery:2008ax}

\bea
\hspace*{-1cm} \lim_{\vect{k}_1+\vect{k}_2 \to 0}\left< \zeta_{\vect{k}_1}\zeta_{\vect{k}_2}\zeta_{\vect{k}_3}\zeta_{\vect{k}_4}\right> &=& \left<\left<\zeta_{\vect{k}_1}\zeta_{\vect{k}_2}\right>_B\left<\zeta_{\vect{k}_3}\zeta_{\vect{k}_4}\right>_B\right>\nonumber\\
&=&(n_s-1)^2 (2\pi)^3\delta^3(\vect{k}_1+\vect{k}_2+\vect{k}_3+\vect{k}_4)P_{\zeta}(k_{12})P_{\zeta}(k_1)P_{\zeta}(k_3)
~,
\eea
with $k_{12}\equiv |\vect{k}_1+\vect{k}_2 |$.

Again from this contribution clearly $\tau_{NL} \sim (\fnl^{(sq)})^2$.
Thus if $\fnl^{(sq)}\sim 50$, the find that there are large contributions to the trispectrum of order $\tau_{NL}\sim 2500$, which are also in an interesting range for Planck, that should be sensitive to a $\tau_{NL} \ga 560$ \cite{Kogo:2006kh}.

\section{Numerical results}
To validate our analytical estimates, and understand in which part of the parameter space the estimates are valid, we have developed a code, that can evaluate numerically the bispectrum for any single field inflation model. In the code we integrate the background evolution and up to three mode functions at a time, to solve for arbitrary momentum triangles. To integrate the different contributions to the bispectrum, we split the integrals in two: To account for the highly oscillating contribution inside the horizon, we use the method of Chen et al \cite{Chen:2008wn}, and partially integrate using the BD-vacuum as an approximation for the mode functions, while to evaluate the integral from typically 4 to 6 e-folds inside the horizon for the lowest wavelength mode until all modes are far outside the horizon, we integrate Eq.~(\ref{eq:bispecintegral}) with the full numerical solution to the mode functions.

To compare the estimates to the numerical solution in a relevant part of the parameter space, we have chosen parameters that neither are ruled out by observations, nor by theoretical considerations, but still generates an interesting signal. As can be seen in Eqs.~(\ref{eq:equiestimate}) and (\ref{eq:ps}), this model has the remarkable property that even though it is a single field model of inflation, it can generate a feature that is readily seen in the bispectrum, but not even with Planck, can be detected in the power spectrum.

In Flauger et al \cite{Flauger:2009ab} different observational and microphysical bounds of the model have been explored. They find that the non-detection of wiggles in the WMAP data limits $\Lambda^4 / \mu^3 \approx \alpha \phi < 10^{-4}$, and that
it is difficult to create realistic microphysical models that have $f < 10^{-4}$. We have therefore chosen a range of models, specified in table \ref{tab:models}, that are characterised by having a currently non-observable wiggle amplitude power spectrum, but an
$\fnl > 5$. For the background potential we use $\mu\sim6 \times 10^{-4}M_{{\rm pl}}$ and the initial value of $\phi\sim11M_{{\rm pl}}$ found in \cite{McAllister:2008hb}, that gives the correct overall amplitude in the power spectrum, and corresponds to approximately 60 e-folds before the end of inflation in the model.

\begin{table}[!ht]\label{tab:models}
\begin{center}
\hspace*{-1cm}\begin{tabular}{@{}c|c|c@{ }c|c|c|c|c|c}
\hline \hline
name & $\mathbf{k}$-config & $k_{\textrm{min}}$ & $k_{\textrm{max\phantom{i}}}$ &
 $f$ & $\alpha$ & $\Lambda$ & $f_{\textrm A}$&$\delta n_s$\cr
\hline
$\alpha_1$ & equilateral & $1$ & $2$ & $3\times10^{-3}$ & $1\times10^{-5}$ & $0$ & $6.50$&$5.63\times10^{-2}$ \cr
$\alpha_2$ & equilateral & $1$ & $2$ & $1\times10^{-3}$ & $7\times10^{-7}$ & $0$ & $7.16$&$6.64\times10^{-3}$ \cr
$\alpha_3$ & equilateral & $1$ & $2$ & $5\times10^{-4}$ & $1\times10^{-7}$ & $0$ & $5.82$&$8.48\times10^{-4}$ \cr
$\alpha_4$ & equilateral & $1$ & $2$ & $1\times10^{-3}$ & $1\times10^{-7}$ & $0$ & $1.02$&$9.60\times10^{-4}$ \cr
$\alpha_5$ & equilateral & $1$ & $2$ & $2\times10^{-3}$ & $4\times10^{-6}$ & $0$ & $7.15$&$2.72\times10^{-2}$ \cr
$\alpha_6$ & equilateral & $1$ & $1.1$ & $1\times10^{-4}$ & $1\times10^{-7}$ & $0$ & $326$& $4.01\times10^{-4}$ \cr
\hline
$\Lambda_1$ & equilateral & $1$ & $2$ & $3\times10^{-3}$ & $0$ & $4\times10^{-4}$ & $6.91$&$5.95\times10^{-2}$\cr
$\Lambda_2$ & equilateral & $1$ & $2$ & $1\times10^{-3}$ & $0$ & $2\times10^{-4}$ & $6.73$&$6.20\times10^{-3}$ \cr
$\Lambda_3$ & equilateral & $1$ & $2$ & $5\times10^{-4}$ & $0$ & $1\times10^{-4}$ & $2.38$&$3.56\times10^{-4}$ \cr
$\Lambda_4$ & equilateral & $1$ & $2$ & $1\times10^{-3}$ & $0$ & $1\times10^{-4}$ & $0.421$&$3.98\times10^{-4}$ \cr
$\Lambda_5$ & equilateral & $1$ & $2$ & $2\times10^{-3}$ & $0$ & $3\times10^{-4}$ & $5.99$&$2.26\times10^{-2}$ \cr
$\Lambda_6$ & equilateral & $1$ & $1.1$ & $1\times10^{-4}$ & $0$ & $1\times10^{-4}$ & $131$&$1.70\times10^{-4}$ \cr
\hline
$\alpha_{\textrm s}$-model & squeezed with $m=1$ & $1$ & $25$ & $3\times10^{-3}$ & $1\times10^{-5}$ & $0$&$-$&$5.63\times10^{-2}$\cr
$\Lambda_{\textrm s}$-model & squeezed with $m=1$ & $1$ & $25$ & $3\times10^{-3}$ & $0$ & $4\times10^{-4}$&$-$&$5.95\times10^{-2}$\cr
\hline
$\alpha_{\textrm{g}}$ & general model & $0.1$ & $1$ & $3\times10^{-3}$ & $1\times10^{-5}$ & $0$&$-$&$5.63\times10^{-2}$\cr
\hline \hline
\end{tabular}
\end{center}
\caption{Numerically evaluated models. All models have $\mu\sim6\times10^{-4}M_{{\rm pl}}$,
and the comoving wave numbers $k$ are measured in units of $(aH)_0 = (aH)_{\phi=11M_{{\rm pl}}}$.}
\end{table}

\begin{figure}
\begin{minipage}[t]{0.47\textwidth}
\begin{center}
\includegraphics[width=\textwidth]{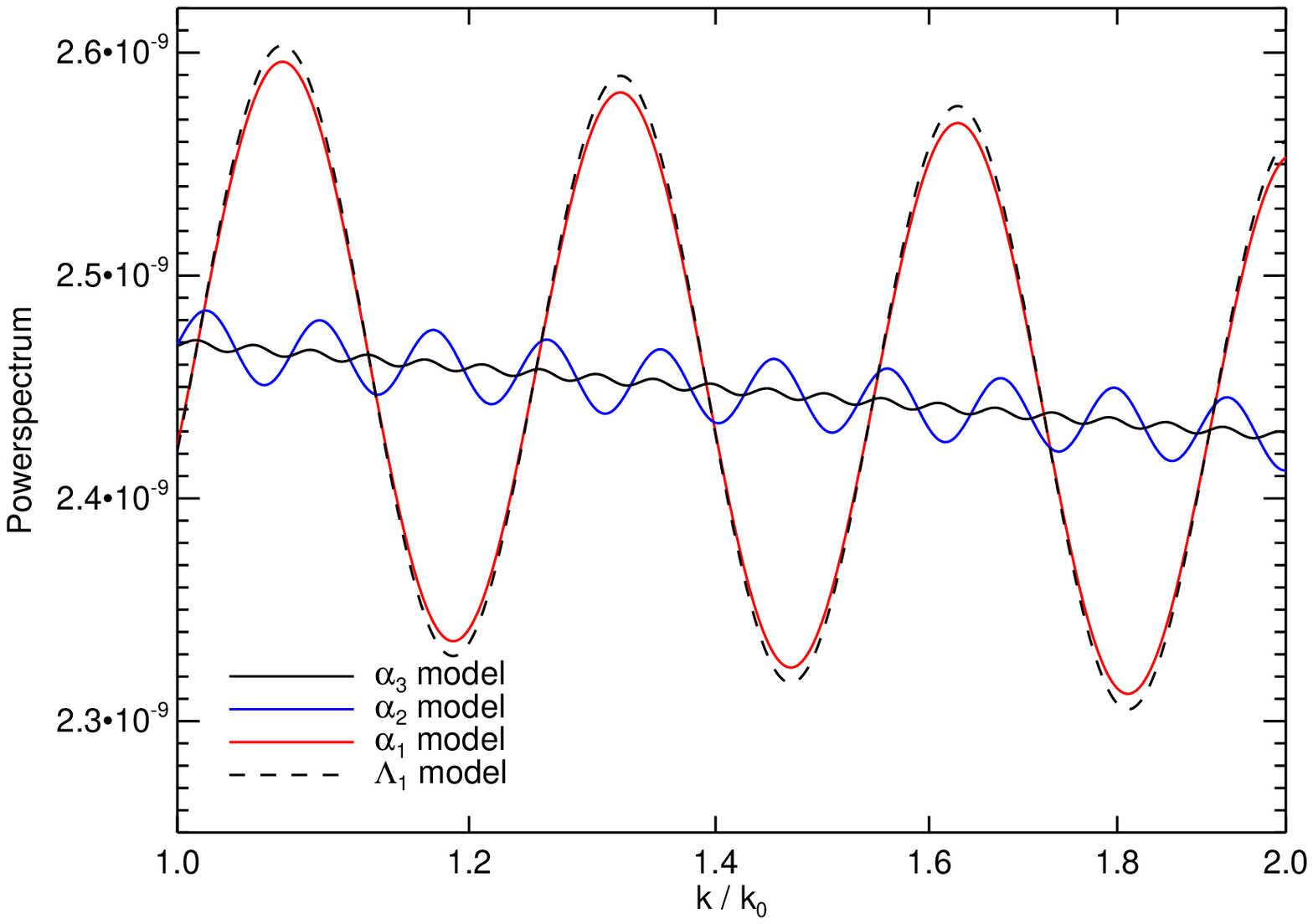}
\end{center}
\end{minipage}
\begin{minipage}[t]{0.47\textwidth}
\begin{center}
\includegraphics[width=\textwidth]{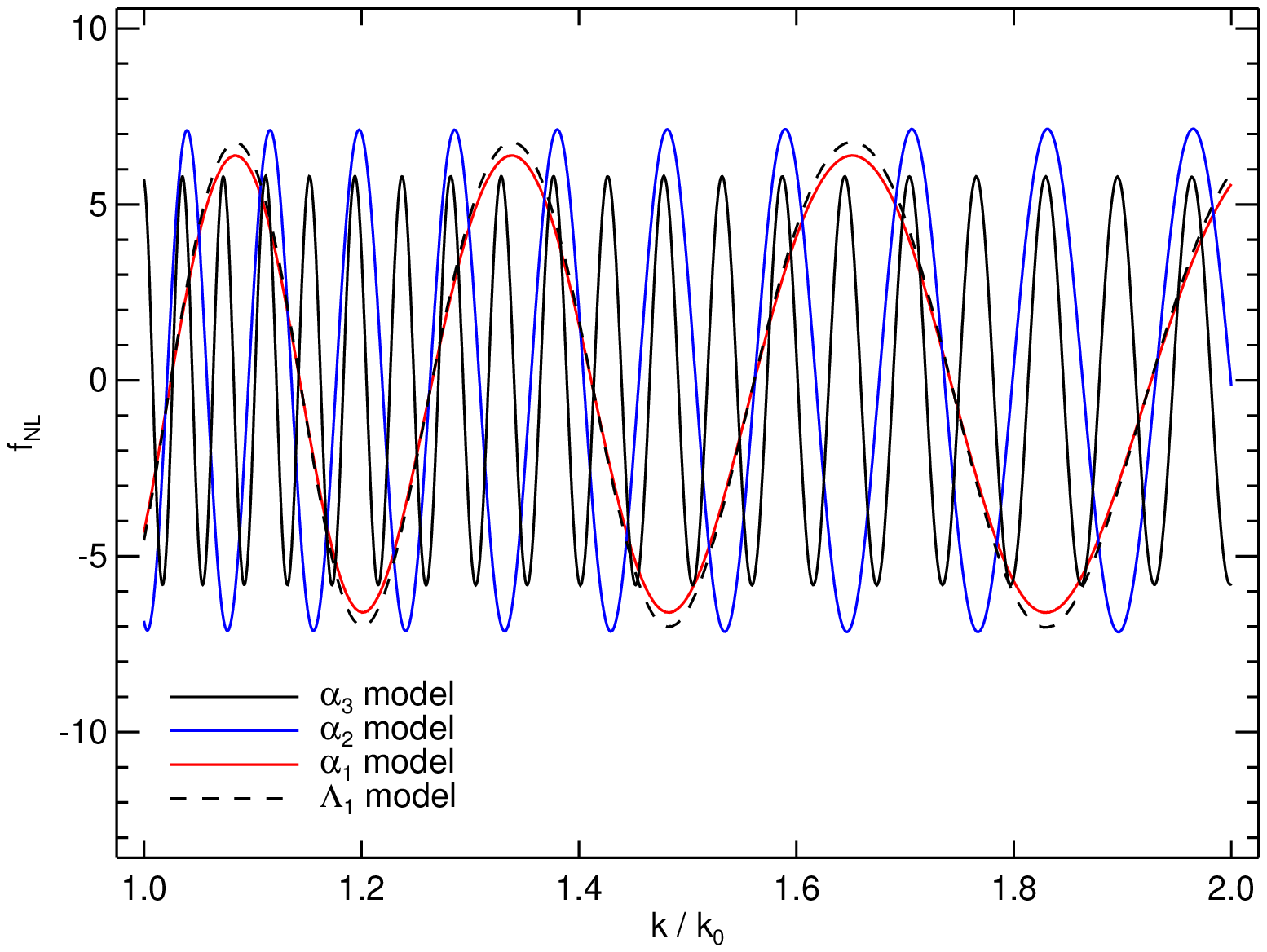}
\end{center}
\end{minipage}
\caption{The power- and bispectra for the models in table}\label{fig:models}
\end{figure}

\begin{figure}
\begin{center}
\includegraphics[width=0.94 \textwidth]{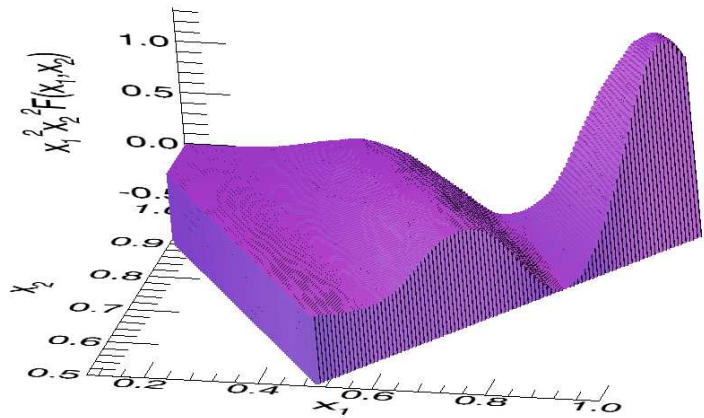}
\caption{The shape function $x_1^2x_2^2F(x_1,x_2)$ as defined in Eq.~(\ref{eq:sf}) for
the general $\alpha_{\textrm{g}}$ model. Note that by definition $F$ is normalised to one for the equilateral form $x_1=x_2=x_3=1$, and that we only plot unique triangles, i.e.~triangles with $x_1>x_2$ have been suppressed in the figure for clarity.}\label{fig:shape}
\end{center}
\end{figure}

\begin{figure}
\begin{center}
\includegraphics[width=0.65 \textwidth]{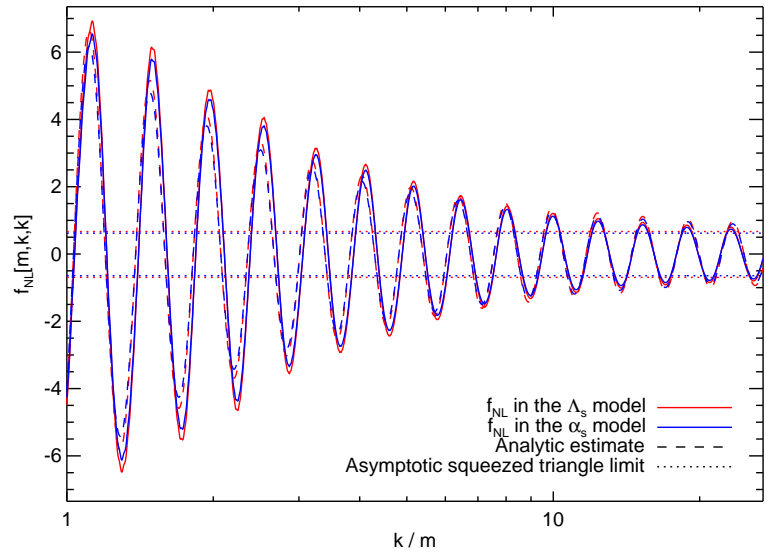}
\caption{The bispectrum for isosceles triangles for the $\alpha_{\textrm s}$ and $\Lambda_{\textrm s}$ models as a function of side ratio $k/m$
together with the analytic model }\label{fig:squeezed}
\end{center}
\end{figure}

\begin{figure}
\begin{center}
\includegraphics[width=0.65 \textwidth]{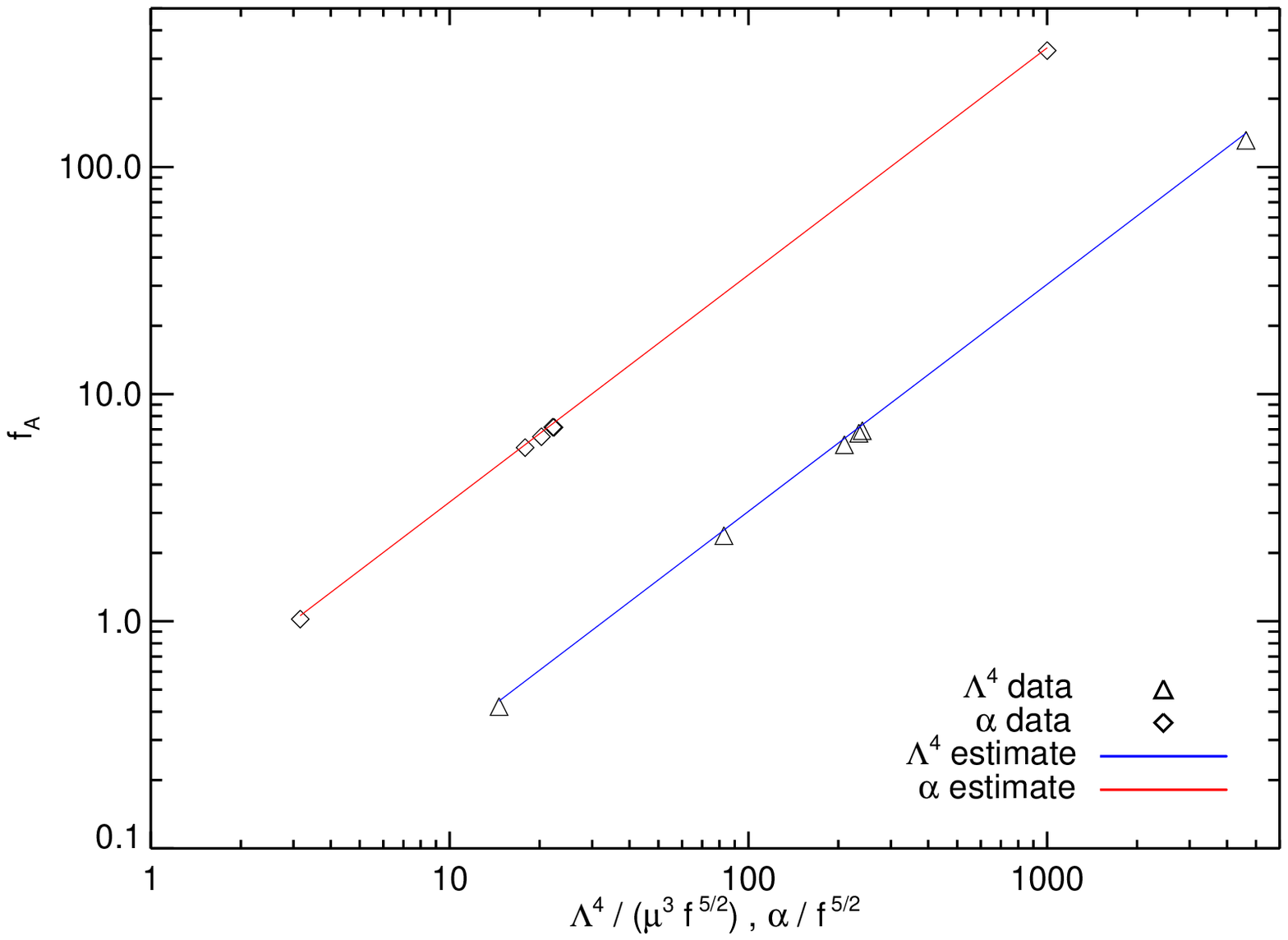}
\caption{Comparison of the analytic estimates of the amplitudes $f_A$ of \fnl~to the numerical results.}\label{fig:fnlestimates}
\end{center}
\end{figure}

The power spectra of some of the models is shown in figure \ref{fig:models}, and it can be seen that for the given values the oscillations in $P_k$ are at the per cent to per mill level. For some of the models the wiggles will not even be detected by the Planck satellite~\cite{Hamann:2008yx}, while at the same time (see figure \ref{fig:models}) they have a significant amount of non-gaussianity.
Ref.~\cite{Hamann:2008yx} studied the detectability of ripples due to trans-Planckian effects with Planck or a future cosmic variance limited experiment. In order for ripples to be unambiguously detected their frequency must lie within a range such that there is a number of oscillations in the observable $k$-range {\it and} their frequency must be lower than the effective width of the experimental window function. With the frequency, $\omega$, defined such that $P(k) \propto \sin(\omega \ln k)$ \cite{Hamann:2008yx} estimates that $1 \la w \la 50$ for it to be detectable. Even if it falls within this range the amplitude must of course also be sufficiently high. At approximately 1$\sigma$ the amplitude, $A$, should be larger than 0.0024 for Planck.
Translating to our case we can make the identification $\omega \sim \phi_* f$ and $A \sim \delta n_s$. To take an example with the $\alpha$-models: $\alpha_3, \alpha_4, \alpha_6$ have too small amplitude to be detected in $P(k)$. $\alpha_1$ and $\alpha_2$ may be detectable, but $\alpha_5$ has $\omega \sim 50$ and may have too fast oscillations for a detection.
In conclusion, a number of the models with a measurably large non-Gaussianity would not have measurable wiggles in the power spectrum.

We have used our code to probe a rather large range of parameter space, and in figure~\ref{fig:fnlestimates} we show how the analytic estimates for the power spectrum and the bispectrum in the equilateral limit are in excellent agreement with the numerical results.
Furthermore, we have also numerically evaluated squeezed triangles (see figure \ref{fig:squeezed}), and found a simple functional form Eq.~(\ref{fnlest}) for general isosceles triangles, which agrees at the 5\%-level with the numerical result.

Given that the bispectrum in this model is nearly scale invariant, it makes sense to characterise the signal in terms of the shape function \cite{Babich:2004gb}
\begin{equation}\label{eq:sf}
F(x_1=\frac{k_1}{k_3},x_2=\frac{k_2}{k_3},1) = \fnl(k_1,k_2,k_3) / \fnl(k_3,k_3,k_3)
\end{equation}
which measures the relative importance of different geometrical configurations compared to the equilateral configuration. In figure \ref{fig:shape} it can be seen how the squeezed configurations are suppressed compared to the equilateral configuration. This is also observed in higher derivative models, and is in contrast to local type non-gaussianity from e.g.~slow-roll inflation, where the squeezed limit is dominating the shape function \cite{Babich:2004gb}. But compared to other models in the literature, this models has the distinct feature of oscillations in \fnl~with an oscillation frequency $\sim \ln K / \phi f$ set by the internal parameters of the theory.

\section{Conclusions}

We have studied in details by numerical and analytical methods the non-Gaussinity generated by a linear single field inflationary potential with superimposed oscillations. This generic form of the potential is an inherent feature of axion monodromy inflation.

The model presents an example of an inflationary model embedded in string theory and can have a very distinct signature on observables. We show that in many cases the main feature of the model is a large non-Gaussianity which is detectable by future experiments such as Planck whereas the impact on the power spectrum is undetectably small. Thus, it appears that the best way to verify this type of models is through a detection of a large bispectrum with the very distinct shape of non-Gaussianity predicted by the model.
However, very interestingly it also shares the usual predictions of chaotic inflation type models, i.e.\ a large tensor-to-scalar ratio and a red spectral index.

Furthermore estimates of the trispectrum in the double-squeezed and in the counter collinear limit shows that non-Gaussianity from the primordial trispectrum could also be detectable by Planck in the present model, although a more careful study of the trispectrum has been left for future studies.

\section*{References} 

\end{document}